\documentclass[10pt,english,american,prl,floatfix,superscriptaddress,twocolumn,showpacs,amsmath,amssymb,reprint]{revtex4-1}
\usepackage[T1]{fontenc}
\usepackage[latin9]{inputenc}
\usepackage{verbatim}
\usepackage{float}
\usepackage{units}
\usepackage{bm}
\usepackage{amsmath}
\usepackage{amssymb}
\usepackage{graphicx}
\usepackage{esint}

\makeatletter

\usepackage[T1]{fontenc}
\PassOptionsToPackage{caption=false}{subfig}
\usepackage{hyperref}
\hypersetup{
breaklinks=true,
colorlinks=true,
citecolor=blue,
linkcolor=blue,
filecolor=blue,
urlcolor=blue
}
\IfFileExists{lmodern.sty}{\usepackage{lmodern}}{}

\makeatother

\usepackage{babel}
\begin{document}
\selectlanguage{english}%

\title{Onset of nonlinear structures due to eigenmode destabilization in
tokamak plasmas}

\author{V. N. Duarte}
\email{vnduarte@if.usp.br, vduarte@pppl.gov}

\selectlanguage{english}%

\address{Institute of Physics, University of São Paulo, São Paulo, SP, 05508-090,
Brazil}

\address{Princeton Plasma Physics Laboratory, Princeton University, Princeton,
NJ, 08543, USA}

\author{H. L. Berk}

\address{Institute for Fusion Studies, University of Texas, Austin, TX, 78712,
USA}

\author{N. N. Gorelenkov}

\address{Princeton Plasma Physics Laboratory, Princeton University, Princeton,
NJ, 08543, USA}

\author{W. W. Heidbrink}

\address{University of California, Irvine, CA, 92697, USA}

\author{G. J. Kramer}

\address{Princeton Plasma Physics Laboratory, Princeton University, Princeton,
NJ, 08543, USA}

\author{D. C. Pace}

\address{General Atomics, San Diego, CA, 92186, USA}

\author{M. Podestà}

\address{Princeton Plasma Physics Laboratory, Princeton University, Princeton,
NJ, 08543, USA}

\author{M. A. Van Zeeland}

\address{General Atomics, San Diego, CA, 92186, USA}

\date{\today}
\selectlanguage{american}%
\begin{abstract}
A general methodology is proposed to differentiate the likelihood
of energetic-particle-driven instabilities to produce frequency chirping
or fixed-frequency oscillations. The method \foreignlanguage{english}{employs
numerically calculated eigenstructures and multiple resonance surfaces
of a given mode in the presence of energetic ion drag and stochasticity
(due to collisions and micro-turbulence). }Toroidicity-induced, reversed-shear
and beta-induced Alfvén-acoustic eigenmodes are used as examples.
\foreignlanguage{english}{Waves measured in experiments are characterized
and compatibility is found between the proposed criterion predictions
and the experimental observation or lack of observation of chirping
behavior of Alfvénic modes in different tokamaks. It is found that
the stochastic diffusion due to micro-turbulence can be the dominant
energetic particle detuning mechanism near the resonances in many
plasma experiments, and its strength is the key as to whether chirping
solutions are likely to arise. The proposed criterion constitutes
a useful predictive tool in assessing whether the nature of the transport
for fast ion losses in fusion devices will be dominated by convective
or diffusive processes.}
\end{abstract}
\maketitle
\selectlanguage{english}%

\section{Introduction}

Supra-thermal fast ions exist in fusion-grade tokamaks as a result
of neutral beam injection (NBI), resonant heating and fusion reactions.
This population of energetic particles (EPs) can strongly resonate
with Alfvénic modes and excite instabilities that can seriously damage
the confinement \cite{Heidbrink2008,Gorelenkov2014,LauberReview2013,ChenZoncaRevModPhys.2016}.
The control of this interaction is necessary for the achievement of
burning plasmas scenarios, in which the fast ions need to have sufficient
time to transfer their energy to the background - mostly from drag
(slowing down) on electrons - in order to ensure high temperature
for the continuation of fusion reactions. This energy transfer mechanism
is considered essential for the good performance of ITER.

Theoretically, the time evolution of the amplitude of a mode interacting
with EPs can exhibit a variety of patterns as the mode departs from
an initial linear phase towards its nonlinear response. During this
evolution, several bifurcations take place, with typical phases being
steady, regular, chaotic and chirping oscillations \cite{BerkPRL1996}.
Upon the kinetic interaction of particles with an eigenmode, nonlinear
phase-space structures may spontaneously emerge in the resonance regions
of the particle distribution, depending on the competition between
drive, damping and collisionality \cite{BerkPLA1997}. These disturbances
can self-consistently support anharmonic oscillations, in a generalization
of BGK modes \cite{BGK1957} which, in the presence of wave damping,
are pushed towards lower energy states. These self-trapped structures
consist of accumulation and depletion of particles in phase-space
and are commonly referred to as clumps and holes, respectively.

Frequency chirping can emerge just above the threshold for marginal
stability where the energy extracted from resonant EPs slightly exceeds
the power being absorbed by the background dissipation, as discussed
in Ref. \cite{BerkPLA1997}. The initial nonlinear response is to
relax the EP distribution in its resonance region, which would reduce
the drive which can then damp the mode. However, the plasma-EP system
can also find an alternate option, of slightly shifting its frequency,
thereby still tapping the free energy of previously untapped neighboring
non-resonant particles, that then become resonant due to the changed
frequency. In the fully developed chirping state, the resonant region
of an enhanced number of particles (clump) or of a deficient number
of particles (holes) are trapped by the finite amplitude wave, and
these regions of phase space shift the resonant distribution to lower
energy regions of phase space, with the released energy being absorbed
by the background dissipation mechanisms that are present while keeping
the nonlinear amplitude hardly changed. Therefore the frequency variation
itself allows for the moving structures to access phase space regions
with distribution function gradients otherwise inaccessible which
leads to \textit{convective} losses over an extended region. 

\selectlanguage{american}%
For the sake of clarity, we distinguish between the terminologies
frequency \textit{chirping} and frequency \textit{sweeping}, that
often appear in the literature. While the former is normally associated
with the fast kinetic response of resonant particles (typically of
order $1ms$) studied in this work, the latter usually corresponds
to a slow evolution (typically $100ms$) of Alfvénic modes, in the
presence of a non-monotonic safety factor $q$ profile, that relies
on the time variation of $q_{min}$. Sweeping events are associated
with a modification of the plasma equilibrium (and consequently, of
the dispersion relation), for example in the case of Alfvén Cascades
\cite{SharapovAlfvenCascadesPoP2001}. Chirping is faster and harder
to suppress using external control.

\begin{figure}[h]
\begin{centering}
\includegraphics[scale=0.6]{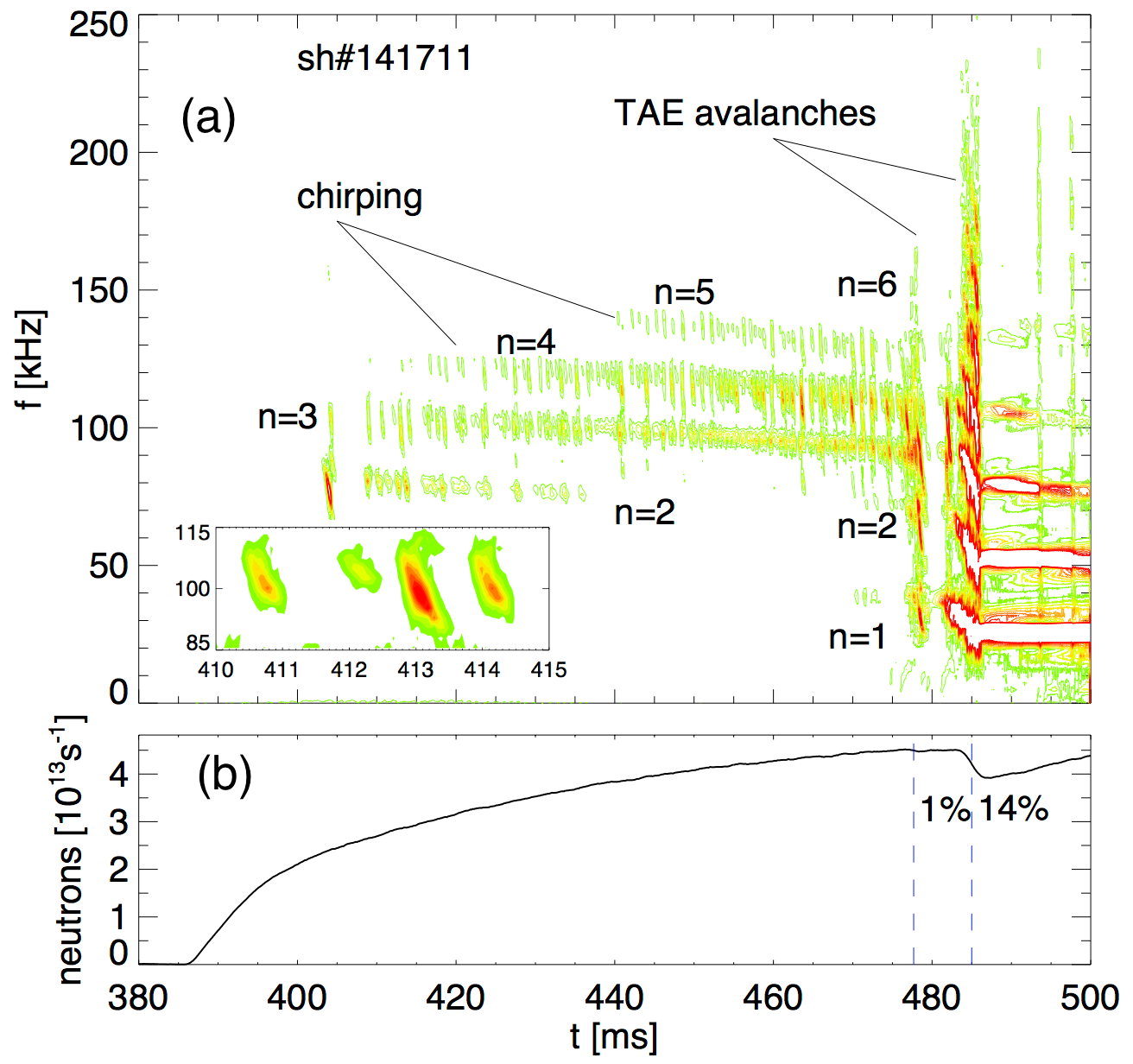}
\par\end{centering}
\caption{(a) Spectrogram showing chirping associated with toroidal Alfvén eigenmodes
(TAEs) for several toroidal mode numbers and (b) neutron loss rates
in NSTX shot 141711 correlating with the TAE avalanches. The small
inset shows a zoomed region with mostly down-chirping.\label{fig:SpectrogramNSTX}}
\end{figure}

\selectlanguage{english}%
Chirping modes can have frequency shifts greater than the linear growth
rate $\gamma_{L}$ and are observed to be precursors of even worst
scenarios, known as avalanches. A spectrogram showing repetitive chirping
cycles followed by avalanches for toroidal Alfvén eigenmodes (TAEs)
in NSTX, for several toroidal mode numbers, is presented in Fig. \ref{fig:SpectrogramNSTX}(a).
The inset shows four of the chirping events and indicates that it
consists mostly of a down-chirping. The system preference for a direction
(up or down in frequency) has been theoretically linked with the competition
between different collisional processes \cite{Lilley2010}. Fig. \ref{fig:SpectrogramNSTX}(b)
shows very significant neutron rate drop correlating with the avalanches. 

The long-range chirping evolution was described by the Berk-Breizman
prediction for the frequency variation $\delta\omega$ scaling with
the bounce frequency $\omega_{b}$ to the power of $3/2$ \cite{BerkPLA1997}.
It has been successfully used for applications that include the inference
of mode amplitude on MAST \cite{Pinches2004} and the estimation of
kinetic parameters such as drive and damping in JT-60U \cite{Lesur2010EspectrDetermination}
and in NSTX \cite{FredricksonPoP2006}.

The present work however focuses on establishing the conditions for
chirping onset rather than modeling their long-term evolution in order
to predict the likely character of EP transport. The EP losses are
typically diffusive (e.g. due to mode overlap, turbulence and collisional
scattering) or convective (as a result of chirping oscillations and
collisional drag). We describe the methodology for the generalization
of previous works and we show how it is possible to include micro-turbulent
stochasticity, which is shown to compete, and even greatly exceed,
collisional scattering in many tokamak experiments and therefore needs
to be added to the stochasticity introduced by pitch-angle scattering. 

Chirping and quasilinear (QL) regimes correspond to two opposite limits
of kinetic theory. Since they may be competing mechanisms in the modification
of the distribution of fast ions in tokamaks (and their consequent
transport), their parameter-space regions of applicability need to
be carefully addressed. The derivation of the QL diffusion equations
\cite{Drummond_Pines_1962,VedenovSagdeev1961} relies on averages,
over a statistical ensemble, that smooth out the distribution function.
In order to justify the resulting smooth, coarse-grained distribution,
stochastic processes (which can be intrinsic due to mode overlap or
extrinsic due to collisions) need to to be invoked. The fast-varying
response associated to the ballistic term is disregarded, which imply
that entropy is no longer conserved. Consequently, QL theory kills
phase correlations and cannot capture chirping events, since chirping
needs time coherence from one bounce to the next in order to move
nonlinear structures altogether over phase space. QL diffusion theory
needs phase decorrelation, i.e. particles need to be expelled from
a phase-space resonant island at a time less than the nonlinear bounce
time. This means that there are no particles effectively trapped.
Due to the reduced dimensionality of phase space, the QL description
is less computationally demanding than the full nonlinear description
needed to capture chirping. It also is much less computationally expensive
than particle codes. A criterion for chirping likelihood is an important
element for identification of parameter space for QL theory applicability
for practical cases and consequent validation of reduced models. An
example of such models is the Resonance-Broadened Quasilinear (RBQ)
code \cite{gorelenkov2016RBQ}. It uses the usual structure of the
QL system written in action-angle variables \cite{KaufmanQLPoF1972}
with a broadened resonance width that scales with bounce frequency,
growth rate and collisional frequency \cite{Berk1995LBQ,GhantousPoP2014}. 

In this work, we build predictive capabilities regarding the likelihood
of the nonlinear regime, which can be useful for burning plasma scenarios.
If further validated and verified, the developed methodologies could
be of practical importance for predictive tools of EP distribution
relaxations in the presence of Alfvénic instabilities. This is especially
important for the development of the reduced models since their methodologies
critically depend on that. 

This paper is organized as follows. In Sec. II the proposed theoretical
methodology is presented. In Sec. III the numerical procedure is presented
along with experimental analysis of results. Discussions are presented
in Sec. IV. and the Appendix is devoted to discussions on the chirping
likelihood in terms of the beam injection energy.

\section{Theoretical framework}

\subsection{Formulation of fast ion interaction with Alfvén waves}

In axisymmetric tokamaks, $\mathcal{E}$, $P_{\varphi}$ and $\mu$
(corresponding respectively to energy, canonical toroidal angular
momentum and magnetic moment) are considered invariants of the unperturbed
motion for EPs interacting with modes that have frequencies much lower
than the cyclotron frequency. Their expressions, all per unit mass
of EPs, are given in S.I. units by

\[
P_{\varphi}\equiv v_{\varphi}R-q_{EP}\psi/m_{EP};
\]
\[
\mathcal{E}=v^{2}/2;
\]
\[
\mu=v_{\perp}^{2}/2B
\]
where $m_{EP}$ and $q_{EP}$ are the mass and charge of EPs, $v$
is the EP speed, $R$ is the tokamak major radius, $\psi$ is the
poloidal flux divided by $2\pi$, $B$ is the magnitude of the magnetic
field, $\theta$ and $\varphi$ are the poloidal and toroidal angles.

If the wave-particle interaction Hamiltonian is assumed, like in the
NOVA code \cite{Gorelenkov1999ChengFu}, to have dependences on time
and poloidal and toroidal angle as $e^{i\left(m\theta-n\varphi-\omega t\right)}$,
where $m$ and $n$ are the wave numbers associated to them, it follows
that upon particle interaction with low-frequency modes, the invariances
of $\mathcal{E}$ and $P_{\varphi}$ are broken but a new invariant
arises: $\mathcal{E}\text{'}=\mathcal{E}+\left(\omega/n\right)P_{\varphi}$
while $\mu$ is kept nearly constant. The existence of two invariants
implies that the resonant particle dynamics is essentially one-dimensional
\cite{BerkPPR1997}. Therefore, a new variable $I=-P_{\varphi}/n$
(at constant $\mathcal{E}'$) can be used to describe this relevant
one-dimensional path for EP dynamics (for steepest distribution function
$f$ modification). The projection of the gradient operator onto this
path is then given by 
\begin{equation}
\frac{\partial}{\partial I}\equiv-n\left.\frac{\partial}{\partial P_{\varphi}}\right|_{\mathcal{E}\text{'}}=-n\frac{\partial}{\partial P_{\varphi}}+\omega\frac{\partial}{\partial\mathcal{E}}.\label{eq:d/dI}
\end{equation}
The resonances for the several harmonics, which are multiple surfaces
in $\left(\mathcal{E},P_{\varphi},\mu\right)$ space, are defined
by

\[
\begin{array}{c}
\Omega_{j}\left(\mathcal{E},P_{\varphi},\mu\right)=\omega+n\left\langle \omega_{\varphi}\left(\mathcal{E},P_{\varphi},\mu\right)\right\rangle \\
-j\left\langle \omega_{\theta}\left(\mathcal{E},P_{\varphi},\mu\right)\right\rangle =0
\end{array}
\]
where $j$ is an integer and $\omega_{\varphi}$ and $\omega_{\varphi}$
are local toroidal and poloidal transit frequencies, respectively.
The phase-space integration is represented as
\[
\begin{array}{c}
\intop d\Gamma...=\int d^{3}r\int d^{3}v...=\\
=\left(2\pi\right)^{3}\underset{\sigma_{\parallel}}{\sum}\int dP_{\varphi}\int d\mathcal{E}/\omega_{\theta}\int m_{EP}cd\mu/q_{EP}...,
\end{array}
\]
where $c$ is the light speed and $\sigma_{\parallel}$ accounts for
counter- and co-passing particles.

\subsection{Collisional operator for fast ions}

Collisional processes are an important element in the determination
of the nonlinear character of wave oscillations \cite{BerkPRL1996,BerkPPR1997,Lilley2009PRL}.
Stochastic processes, such as pitch-angle scattering, act to destroy
the coherent structures that support wave chirping while drag, or
slowing down, is formally equivalent to chirping and enhance the convective
transport of these nonlinear structures. For EPs, the Fokker-Planck
collisional operator that enters the kinetic equation can be approximately
written as a superposition of pitch-angle scattering and drag as follows
\cite{GOLDSTON198161},

\[
C[f]=\frac{\partial}{\partial{\bf v}}\cdot\left[\nu_{\perp}\left(v^{2}\mathbb{I}-{\bf v}{\bf v}\right)\right]\cdot\frac{\partial f}{\partial{\bf v}}+\frac{1}{\tau_{s}}\frac{\partial}{\partial{\bf v}}\cdot\left[{\bf v}\left(1+\frac{v_{c}^{3}}{v^{3}}\right)f\right]
\]
where $f$ is the distribution function; $\nu_{\perp}$ and $\tau_{s}^{-1}$
are the $90\textdegree$ pitch angle scattering rate and the inverse
slowing down time, which are given by
\begin{equation}
\nu_{\perp}=\frac{1}{2}\left\langle Z\right\rangle \frac{\overline{A}_{i}}{\left[Z\right]}\frac{1}{A_{EP}}\left(\frac{v_{c}}{v}\right)^{3}\frac{1}{\tau_{s}}\label{eq:nu_perp}
\end{equation}

\begin{equation}
\tau_{s}^{-1}=\frac{Z_{EP}^{2}e^{4}m_{e}^{1/2}n_{e}ln\Lambda_{e}}{3\left(2\pi\right)^{3/2}\epsilon_{0}^{2}m_{EP}T_{e}^{3/2}}\label{eq:1/tau_S}
\end{equation}
The critical speed, above which electron drag dominates over ion drag,
is defined as

\begin{equation}
v_{c}=\left(\frac{3\sqrt{\pi}}{4}\frac{m_{e}}{m_{p}}\frac{\left[Z\right]}{\overline{A}_{i}}\right)^{1/3}\sqrt{\frac{2T_{e}}{m_{e}}}\label{eq:v_crit}
\end{equation}
and
\[
\frac{\left[Z\right]}{\overline{A}_{i}}=\underset{\beta}{\sum}\frac{Z_{\beta}^{2}n_{\beta}}{A_{\beta}n_{e}}\frac{ln\Lambda_{\beta}}{ln\Lambda_{e}}
\]

\[
\left\langle Z\right\rangle =\underset{\beta}{\sum}\frac{Z_{\beta}^{2}n_{\beta}}{n_{e}}\frac{ln\Lambda_{\beta}}{ln\Lambda_{e}}
\]
For a given ion species $\beta$, $A_{\beta}=m_{\beta}/m_{p}$, with
$m_{p}$ being the proton mass, $ln\Lambda$ is the Coulomb logarithm,
$Z_{i}$ and $Z_{EP}$ are the background and energetic ion atomic
number, $n$ is the density, $T$ is the temperature, $m$ is the
mass and the thermal speed for a given species \textit{a} is defined
as $v_{Ta}=\sqrt{2T_{a}/m_{a}}$. The three-dimensional velocity derivatives
are projected onto the aforementioned path of the steepest gradient
of the distribution function so that
\[
\frac{\partial}{\partial{\bf v}}\rightarrow\left.\frac{\partial\Omega}{\partial P_{\varphi}}\right|_{\mathcal{E}\text{'}}\frac{\partial P_{\varphi}}{\partial{\bf v}}\frac{\partial}{\partial\Omega}
\]
Since $\frac{\partial P_{\varphi}}{\partial{\bf v}}=R\hat{\varphi}$
and $v_{\varphi}\thickapprox v_{\parallel}\frac{B_{\varphi}}{B}$,
it follows that the effective collisional operator for fast ions resonating
with a given mode can be approximately cast in the form

\[
C[f]=\nu_{scatt}^{3}\frac{\partial^{2}f}{\partial\Omega^{2}}+\nu_{drag}^{2}\frac{\partial f}{\partial\Omega}
\]
where the effective scattering $\nu_{scatt}$ and drag $\nu_{drag}$
coefficients are

\begin{equation}
\nu_{scatt}^{3}\simeq2\nu_{\perp}R^{2}\left[\mathcal{E}-\frac{B_{\varphi}^{2}}{B^{2}}\left(\mathcal{E}-\mu B\right)\right]\left(\left.\frac{\partial\Omega}{\partial P_{\varphi}}\right|_{\mathcal{E}\text{'}}\right)^{2}\label{eq:scatt,eff}
\end{equation}
and 
\begin{equation}
\nu_{drag}^{2}\simeq\frac{\sqrt{2}R}{\tau_{s}}\sqrt{\mathcal{E}-\mu B}\frac{B_{\varphi}}{B}\left(1+\frac{v_{c}^{3}}{v^{3}}\right)\left.\frac{\partial\Omega}{\partial P_{\varphi}}\right|_{\mathcal{E}\text{'}}\label{eq:drag,eff}
\end{equation}
In order to be properly evaluated, the above expressions need to be
averaged over the orbit bounce motion of a particle, as detailed in
the section \ref{subsec:Averaging-implementations-in}. 

\subsection{Inclusion of micro-turbulent stochasticity}

Stochastic diffusion is determined by collisional scattering processes,
such as pitch angle scattering, as well as additional processes, such
as the effect of the background turbulence that is often dominant
in the determination of the global heat outflow. Micro-turbulence
is introduced using a procedure that follows the one introduced by
Lang and Fu \cite{LangFu2011}, where it is considered that the main
contribution comes from electrostatic ion temperature gradient (ITG)
turbulence via radial diffusion rather than the velocity diffusion.
As with collisional scattering, the turbulent diffusion operator is
projected onto the relevant one-dimensional path for particle dynamics,
represented by the variable $\Omega$. Since $\frac{\partial}{\partial r}=\left.\frac{\partial\Omega}{\partial P_{\varphi}}\right|_{\mathcal{E}\text{'}}\frac{\partial P_{\varphi}}{\partial r}\frac{\partial}{\partial\Omega}$
with $\frac{\partial P_{\varphi}}{\partial r}\approx-\frac{q_{EP}}{m_{EP}}\frac{\partial\psi}{\partial r}$,
the spatial micro-turbulence diffusion operator along the radial coordinate
$r$ can be written as

\begin{equation}
\frac{1}{r}\frac{\partial}{\partial r}rD_{EP}\frac{\partial f}{\partial r}\approx D_{EP}\left(\frac{q_{EP}}{m_{EP}}\frac{\partial\psi}{\partial r}\right)^{2}\left(\left.\frac{\partial\Omega}{\partial P_{\varphi}}\right|_{\mathcal{E}\text{'}}\right)^{2}\frac{\partial^{2}f}{\partial\Omega^{2}}\label{eq:RadialDiffOp}
\end{equation}
where $D_{EP}$ is the EP diffusivity. Therefore, the ratio between
the effective stochasticities coming from micro-turbulent (Eq. \ref{eq:RadialDiffOp})
and collisional processes (Eq. \ref{eq:scatt,eff}) is

\begin{equation}
R\approx\frac{D_{EP}\left(\frac{q_{EP}}{m_{EP}}\frac{\partial\psi}{\partial r}\right)^{2}}{2\nu_{\perp}R^{2}\left[\mathcal{E}-\frac{B_{\varphi}^{2}}{B^{2}}\left(\mathcal{E}-\mu B\right)\right]}\label{eq:RatioTurbScatt}
\end{equation}
A subtlety in applying the model relies on the determination of $D_{EP}$.
It is expected to be lower than thermal ion diffusivity $D_{th,i}$
since the micro-turbulence wavelength is typically smaller than the
beam cyclotron orbit. \foreignlanguage{american}{Historically the
effect of micro-turbulence on EP transport has been neglected based
on the fact that since EPs have large orbits, they should experience
several phases of the turbulent fields in such a way as to cancel
out its overall effect. Although fast ions turbulent transport is
negligible compared to Alfvénic-induced transport (see, for example,
studies on ASDEX-U \cite{Geiger2015PPCF} and DIII-D \cite{Pace2013}),
turbulence can be an important transport mechanism, as compared to
collisions, at the onset of the evolution of modes, when their amplitude
are still small.} Over the past decade, the modeling of $D_{EP}$
has been studied by several groups \cite{Hauff2009PRL,PueschelNF2012,EstradaMila2006,Albergante2010,Albergante2009PoP,ZhangLinChen2008PRL}.
We have chosen to determine $D_{EP}$ from the scalings that follow
from gyrokinetic simulations of GTC \foreignlanguage{american}{\cite{ZhangLinChen2008PRL},
in which $D_{EP}$ is proportional to the thermal ion diffusivity,
$D_{i}$, and a function of $T_{i}/E_{EP}$, where $E_{EP}$ is the
energy of the EPs. The GTC simulations assumed a specific plasma background,
that can be considerably different from a given discharge being analyzed.
Therefore, a significant error can be expected to be associated to
the inferred value of $D_{EP}$. Alternative ways of obtaining $D_{EP}$
could be achieved by using Refs. \cite{PueschelNF2012,Hauff2009PRL}.
In our analysis we infer the value of $D_{i}$ from the outputs of
the global transport code TRANSP \cite{Hawryluk1980,GOLDSTON198161}
at the position where the mode structure is peaked at the time being
analyzed. The particle diffusivity is known to have a huge error associated
to it because TRANSP cannot resolve well particle sources, especially
close to wall. On the other hand, the thermal conductivity $\chi$
is reasonably well known and therefore is frequently used \cite{Heidbrink2009PRL}
as an indication of the actual value of $D$ (the exact relation for
a Maxwellian distribution would be $D=2\chi/3$). }

\subsection{The early nonlinear evolution of a mode amplitude }

The onset of a mode amplitude evolution can be studied using perturbation
theory within the kinetic framework. Refs. \cite{BerkPRL1996,BreizmanPoP1997,BerkPPR1997,Lilley2009PRL}
showed that, for $\left|\gamma_{L}-\gamma_{d}\right|\ll\gamma_{L}$,
truncation of mode amplitude at third order is justified due to fact
that the unstable system is close to marginal stability. Taking $\nu_{stoch}$
(the overall stochasticity felt by EPs, which includes $\nu_{scatt}$)
and $\nu_{drag}$ independent of time but dependent on phase space
localization, the equation for the early-time perturbed mode (a mode
that exists without accounting for the kinetic component) amplitude
evolution can be written as a time-delayed, integro-differential cubic
equation 

\begin{equation}
\begin{array}{c}
\frac{dA(t)}{dt}-A(t)=-\underset{j}{\sum}\intop d\Gamma\mathcal{H}\int_{0}^{t/2}d\tau\tau^{2}A\left(t-\tau\right)\times\\
\times\int_{0}^{t-2\tau}d\tau_{1}e^{-\hat{\nu}_{stoch}^{3}\tau^{2}\left(2\tau/3+\tau_{1}\right)+i\hat{\nu}_{drag}^{2}\tau\left(\tau+\tau_{1}\right)}\times\\
\times A\left(t-\tau-\tau_{1}\right)A^{*}\left(t-2\tau-\tau_{1}\right)
\end{array}\label{eq:cubic-1}
\end{equation}
where $\mathcal{H}=2\pi\omega\delta\left(\Omega_{j}\right)\left|V_{n,j}\right|^{4}\left(\frac{\partial\Omega_{j}}{\partial I}\right)^{3}\frac{\partial f}{\partial\Omega}$
and 
\[
V_{l_{1},l_{2},l_{3}}\left(I\right)=\frac{iq_{EP}}{\omega}\int\frac{d\xi_{1}d\xi_{2}d\xi_{3}}{\left(2\pi\right)^{3}}e^{-i(l_{1}\xi_{1}+l_{2}\xi_{2}+l_{3}\xi_{3})}{\bf v\cdot e}
\]
or
\begin{equation}
V_{n,j}\left(I\right)=\frac{iq_{EP}}{\omega}\int\frac{d\varphi d\theta}{\left(2\pi\right)^{2}}e^{-i(j\theta-n\varphi)}{\bf v\cdot e}\label{eq:V_l}
\end{equation}

accounts for the wave-particle energy exchange, where $\mathbf{e}$
is the electric field eigenstructure and $\mathbf{v}$ is the velocity
of a resonant particle. $l_{1},l_{2},l_{3}$ are integers and $\xi_{1},\xi_{2},\xi_{3}$
are angles conjugated to the invariants of motion (actions of the
Hamiltonian). In equation \eqref{eq:cubic-1}, the circumflex denotes
normalization with respect to $\gamma=\gamma_{L}-\gamma_{d}$ (growth
rate minus damping rate) and $t$ is the time normalized with the
same quantity. $A$ is the normalized complex mode amplitude of an
eigenmode oscillating with frequency $\omega$.

The solutions of eq. \eqref{eq:cubic-1} can exhibit several bifurcations
and therefore several phases, as shown for a bump-on-tail configuration
in Ref. \cite{BerkPRL1996}. Interestingly, eq. \eqref{eq:cubic-1}
allows for the excitation of sub-critical instabilities and for nonlinear
frequency splitting \cite{FasoliPRL1998}. If the nonlinearity in
equation \eqref{eq:cubic-1} is weak, the system will most likely
saturate close to the linear stability state, where the trapping frequency
$\omega_{b}$ satisfies $\omega_{b}\ll\gamma_{L}\simeq\gamma_{d}$.
However, in case the solution of the cubic equation explodes, the
system enters a strong nonlinear phase, which may lead to chirping
modes. Indeed, long-range numerical simulations indicate the explosive
behavior of $A$ as the precursor to the formation of holes and clumps
structures \cite{BerkCandy1999}. Furthermore, chirping events are
significantly enhanced by the coherence introduced by dynamical friction
(i.e, particle drag) \cite{Lilley2009PRL,Lilley2010} and are inhibited
by stochasticity from diffusive processes, such as resonant particle
heating \cite{MaslovskyMauelPRL2003}, collisional pitch-angle scattering
and from background turbulence, all of which contribute to causing
particles to detune from a resonance. Stochastic events lead to loss
of phase information that contribute to destroy coherent structures. 

Eq. \eqref{eq:cubic-1} was originally derived for a bump-on-tail
system with Krook collisions \cite{BerkPRL1996} and later generalized
to complex tokamak geometries and also to include collisional scattering
\cite{BerkPPR1997}. Lilley, Breizman and Sharapov \cite{Lilley2009PRL}
included the effects of drag on the bump-on-tail cubic equation and
derived a criterion to determine stable and unstable regions of solutions
of the cubic equation in drag \textit{vs} scattering collisional parameter
space. Our work aims at improving this prediction by using realistic
resonant surfaces and mode structure information, coming from the
NOVA and NOVA-K codes. To this end, orbit and phase space averages
are employed in order to account for the effective Fokker-Planck collisional
coefficients. Experimental data are analyzed in order to verify whether
chirping events coincide with the occurrence with the \textquotedbl{}explosive\textquotedbl{}
phase of the cubic equation, as predicted by the theory. 

\subsection{A criterion for chirping onset}

It has been shown \cite{DuarteAxivPRL} that a simplified bump-on-tail
approach that only accounts for a single representative value for
the collisional coefficients is insufficient to make predictions for
a mode nonlinear nature in tokamaks. The missing physics were shown
to be the absence of non-uniform mode structures, (multiple) resonance
surfaces and poloidal bounce averages that account for particle trajectories
on a poloidal cross section.

\begin{figure}[H]
\includegraphics[scale=0.35]{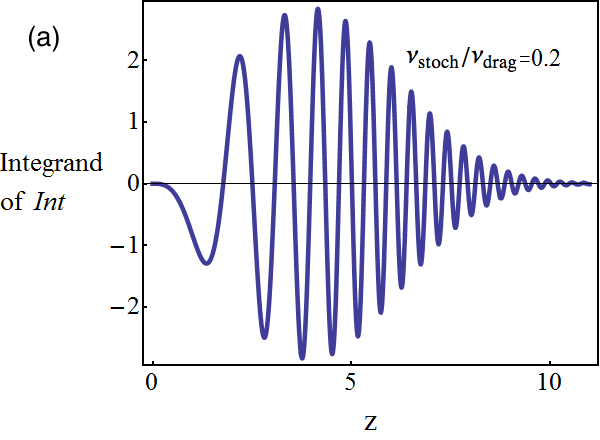}

\includegraphics[scale=0.35]{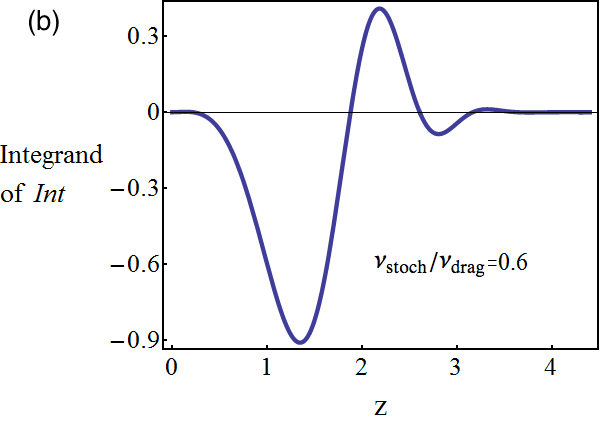}

\includegraphics[scale=0.35]{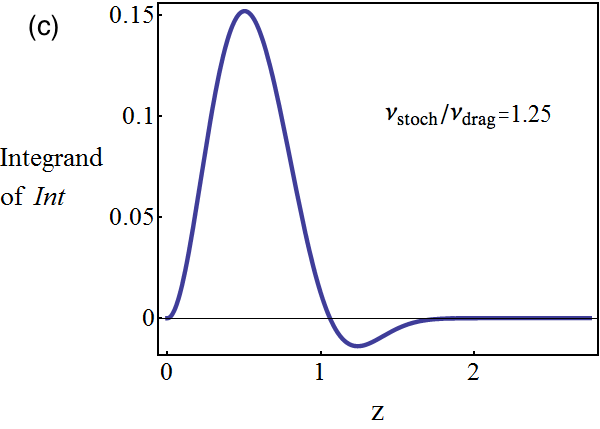}

\caption{Integrand of $Int,$ given by eq. \eqref{eq:Int-1} as a function
of $\nu_{stoch}/\nu_{drag}$. It has a strong oscillating behavior
for small values of $\nu_{stoch}/\nu_{drag}$, which makes evident
the changing sign introduced by drag in the kernel of the cubic equation
\eqref{eq:cubic-1} (part a). In this regime the sign of the integral
flips recurrently and prevents a steady solution to being established.
For moderately higher values of $\nu_{stoch}/\nu_{drag}$, the integrand
is less oscillatory but the integral is still negative (part b). After
$\nu_{stoch}/\nu_{drag}$ exceeds $1.04$, the integral becomes positive.
Part c was taken close to the peak positive value of the integral.\label{fig:IntIntegrands}}
\end{figure}

A necessary but not sufficient condition for the existence of fixed-frequency,
steady-state solutions would be that the real component of the right-hand
side of Eq. \eqref{eq:cubic-1} be negative at late times when the
response is stationary, i.e. when the nonlinear term is allowed to
balance the linear growth. The delta function $\delta\left(\Omega_{j}\left(P_{\varphi},\mathcal{E},\mu\right)\right)$
that restricts the integration to the resonance condition can be exploited
and the following criterion for the existence of fixed-frequency oscillations
is obtained \cite{DuarteAxivPRL}:
\begin{equation}
Crt=\frac{1}{N}\underset{j,\sigma_{\parallel}}{\sum}\int dP_{\varphi}\int d\mu\frac{\left|V_{n,j}\right|^{4}}{\omega_{\theta}\nu_{drag}^{4}}\left|\frac{\partial\Omega_{j}}{\partial I}\right|\frac{\partial f}{\partial I}Int>0\label{eq:Criterion-1}
\end{equation}
where
\begin{equation}
Int\equiv Re\int_{0}^{\infty}dz\frac{z}{\frac{\nu_{stoch}^{3}}{\nu_{drag}^{3}}z-i}exp\left[-\frac{2}{3}\frac{\nu_{stoch}^{3}}{\nu_{drag}^{3}}z^{3}+iz^{2}\right]\label{eq:Int-1}
\end{equation}
and $N$ is a normaliation for $Crt$, which consists in the same
sums and integrations that appear in the numerator of \eqref{eq:Criterion-1}
but in the absence of $Int$. In eqs. \eqref{eq:Criterion-1} and
\eqref{eq:Int-1} , the quantities $\tau_{b}$, $\nu_{drag}$, $\nu_{stoch}$,
$V_{n,j}$ and $\Omega_{j}$ are understood to be evaluated at $\mathcal{E}=\mathcal{E}'-\omega P_{\varphi}/n$.
Criterion \eqref{eq:Criterion-1} was incorporated into NOVA-K making
use of a polynomial interpolation for $Int$. $Crt$ provides a prediction
for the likelihood of a fully nonlinear phenomenon obtained only from
pure linear physics elements and therefore can be tested in linear
codes. This is a considerable advantage in efficiency for making a
prediction of a nonlinear property. The integrand of $Int$ is plotted
in figure \eqref{fig:IntIntegrands} for different values of $\nu_{stoch}/\nu_{drag}$. 

Drag introduces an oscillatory behavior of the integrand of $Int$.
This has an effect of flipping the sign of the integral kernel of
the cubic equation. This causes $dA/dt$ to vary more abruptly and
therefore prevents a steady state solution from being achieved. Instead
nonlinear chirping solution is more likely to be achieved. It is interesting
to note that formally, drag enters the kinetic equation in a mathematically
similar way as a chirping frequency does, with $kv\nu_{drag}$ replaced
by $d\omega/dt$.

\begin{figure}[H]
\includegraphics[scale=0.35]{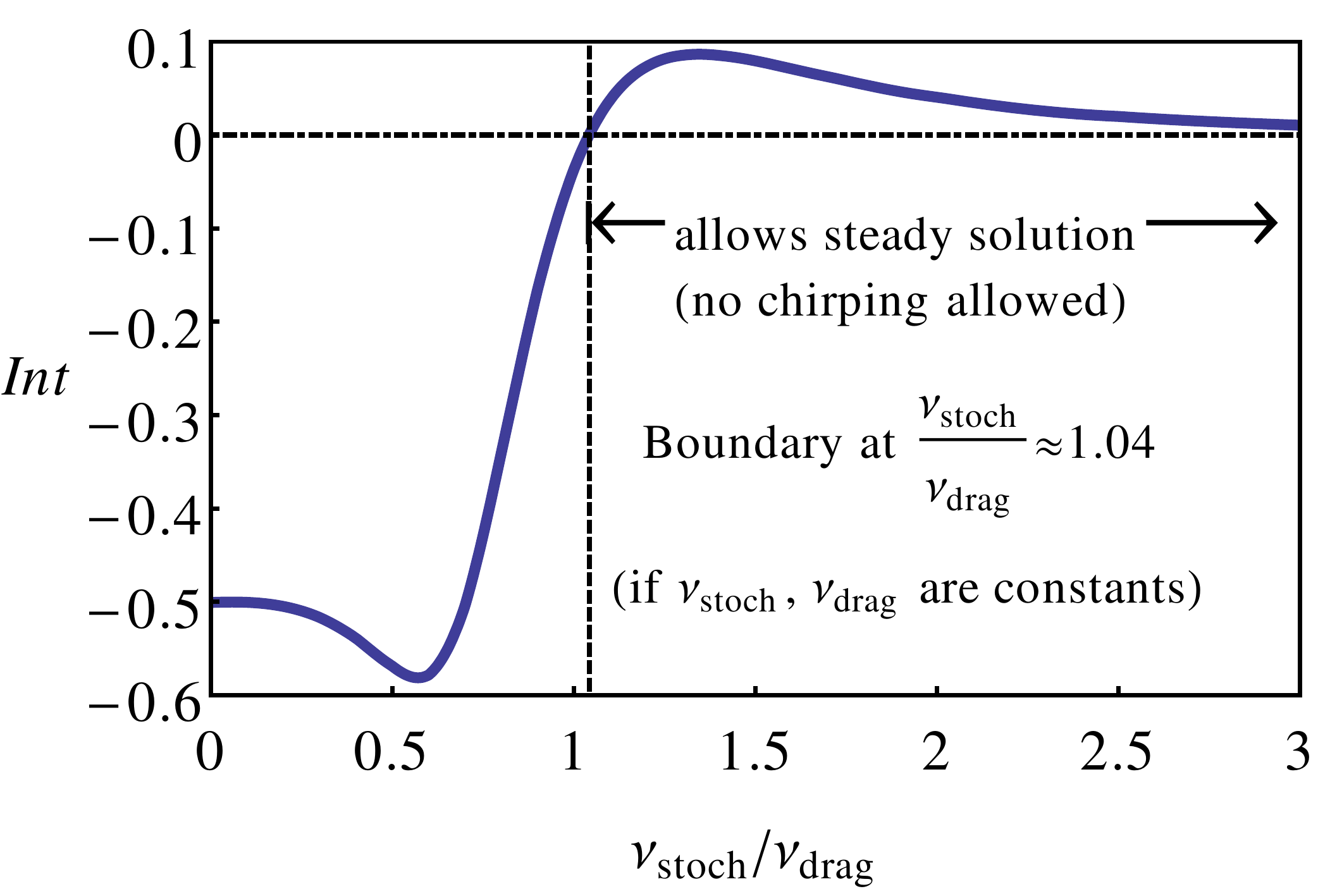}

\caption{Numerical values for $Int$, the time integral in equation \eqref{eq:Int-1},
as a function of $\nu_{stoch}/\nu_{drag}$. An interpolation of this
function is provided as an input to NOVA-K code, which performs the
phase space integrals \textit{a posteriori}. \label{fig:Int}}
\end{figure}

$Int$, which is a function of phase space, is plotted in Fig. \eqref{fig:Int}
as a function of $\nu_{stoch}/\nu_{drag}$. Note that the positive
and negative domains of $Int$ are roughly an order of magnitude different
in the interval $0\leq\nu_{stoch}/\nu_{drag}\lesssim2$, while for
$\nu_{stoch}/\nu_{drag}\gg1$, $Int\approx1.022\left(\nu_{stoch}/\nu_{drag}\right)^{-4}$. 

For our investigation, we focus on the criterion for the non-existence
of a fixed-frequency, steady solution. To determine the physical parameters
needed in the calculation described below, the global transport code
TRANSP \cite{Hawryluk1980,GOLDSTON198161} is used. The code determines
the needed particle diffusivity that matches the observed plasma parameters
to the particle and heat sources. We interpret this empirically obtained
result as due to both classical (which includes neoclassical) transport
and due to micro-turbulence. The resulting thermal particle diffusivity
is then used together with a model to scale the EP diffusivity given
the turbulent background diffusivity, to estimate the diffusivity
of EPs, which will then contribute to the value of $\nu_{stoch}$.
Here we perform a quantitative study based on the time delayed cubic
equation using experimental results to determine whether the theoretically
predicted nonlinear character of the response correlates with the
observation. 

\section{Numerical study of the chirping criterion}

\subsection{Kinetic-MHD perturbative computations}

NOVA \cite{Gorelenkov1999ChengFu} is a nonvariational, hybrid kinetic-MHD
stability code primarily used to integrate non-Hermitian integro-differential
eigenmode equations in the presence of EPs, using a general flux coordinate
system. It uses realistic numerically calculated MHD equilibria and
hence it is suited to study both conventional and spherical tokamaks.
The code uses Fourier expansion in $\theta$ and cubic spline finite
elements in the radial $\psi$ direction. NOVA provides the eigenstructures
used in our analysis.

NOVA-K \cite{CHENG1992,Gorelenkov1999Saturation} takes into account
finite Larmor radius and orbit width efects to study the destabilization
of MHD modes from the EPs free energy. We use NOVA-K to calculate
the resonance surfaces, in $\left(P_{\varphi},\mathcal{E},\mu\right)$
space, associated with the modes. These surfaces are needed to calculate
the growth and damping rates of eigenmodes in the presence of EPs
as well as the integral $Int$. TRANSP is used to provide the distribution
function which contains the necessary information on the most representative
EP population as input for NOVA-K runs. 

The quadratic form of MHD is particularly useful when stability analysis
is addressed. If the mode frequency is assumed to be $\omega+i\gamma$,
with $\left|\gamma\right|\ll\left|\omega\right|$ where $\omega$
is the mode eigenfrequency and $\gamma$ is its growth rate, it is
obtained \cite{Gorelenkov1999ChengFu,CHENG1992}
\[
\gamma\approx\frac{Im\delta W_{k}}{2\delta K}\omega
\]
where $\delta K=\omega^{2}\int\rho\left|\bm{\xi}\right|^{2}d{\bf r}/2$
is the inertial (kinetic) energy, $\delta W_{k}$ is the potential
energy associated to the nonadiabatic component of the distribution
function and $\omega\delta W_{k}$ is the power released by the resonant
particles. The growth rate can be expressed as 

\begin{equation}
\begin{array}{c}
\gamma=\frac{2M^{2}\pi^{3}c}{q_{EP}\int\rho\left|\bm{\xi}\right|^{2}d{\bf r}}\underset{\sigma_{\parallel}}{\sum}\int dP_{\varphi}d\mu d\mathcal{E}\underset{m,m',j}{\sum}G_{m'j}^{*}\mathcal{E}^{2}\tau_{b}\\
\left(\frac{\partial f}{\partial\mathcal{E}}\right)\left(1-\nicefrac{\omega_{*}}{\omega}\right)G_{mj}\delta\left(\omega+n\omega_{\varphi}-j\omega_{\theta}\right)
\end{array}\label{eq:gammaNOVA}
\end{equation}
\foreignlanguage{american}{with the diamagnetic frequency being defined
by}

\[
\omega_{*}=-i\frac{\frac{\partial f}{\partial P_{\varphi}}}{\frac{\partial f}{\partial\mathcal{E}}}\frac{\partial}{\partial\varphi}
\]
$G$ represents mode structure matrix elements, which are associated
to the projection of the resonant particle current onto the wave electric
field. For each isolated resonance, the reduced Hamiltionian can be
written as \cite{BerkPPR1997}

\[
\begin{array}{c}
H\left(I,\zeta,t\right)=H_{0}(I)+H_{1}\left(I,\zeta,t\right);\;\\
H_{1}\left(I,\zeta,t\right)=2A(t)V_{n,j}(I)cos\left(\zeta-\omega t\right)
\end{array}
\]
where $\zeta\equiv l_{1}\xi_{1}+l_{2}\xi_{2}+l_{3}\xi_{3}$ and the
bounce frequency of the most deeply trapped particles can be shown
to be

\[
\omega_{b}=\left|2A(t)V_{n,j}(I_{r})\left.\frac{\partial\Omega_{j}}{\partial I}\right|_{I=I_{r}}\right|^{1/2}
\]
where the subscript $r$ denotes the resonance location and the derivative
is defined by eq. \eqref{eq:d/dI}. The $G$ matrices are related
to $V$ (given by equation \eqref{eq:V_l}) via

\selectlanguage{american}%
\[
\left(\frac{M\mathcal{E}}{\sqrt{2}}\right)^{2}\sum_{m,m'}G_{m'j}^{*}G_{mj}=\left|V_{n,j}\right|^{2}
\]

\selectlanguage{english}%

\subsection{Mode structure identification}

In order to characterize the mode being observed in the experiment,
NSTX and TFTR reflectometer measurements are compared to the mode
structures computed by NOVA. Reflectometry diagnostic measures the
density fluctuation of the plasma at the location where the launched
wave has a cutoff. The fluid displacement times the local density
gradient is equivalent to the density fluctuation. All poloidal harmonics
calculated by NOVA are summed up for this analysis. Modes are categorized
according to their mode structure and whether their frequencies fall
into a given gap of the continuum. The reflectometer cannot resolve
the density in the core for the cases of flat density profile and
especially when the peak of the density is displaced from the plasma
center, as typically observed during the H-mode regime. In DIII-D,
Electron Cyclotron Emission is used for the purpose of measuring the
mode structure, following a methodology described in \cite{VanZeeland2006}.
An interesting aspect of the observation is that, in spite of the
intrinsic nonlinear nature of chirping events, the structure of the
chirping mode is not substantially changed during the time evolution
of the system \cite{FredricksonPoP2006,Podest2012,Heidbrink1995Chirping}.
This gives credence to the use of a linear code for the eigenstructure
identification at early times.

\subsection{\label{subsec:Averaging-implementations-in}Averaging implementations
in NOVA-K }

In order to calculate the above expressions for the effective collisional
coefficients \eqref{eq:scatt,eff} and \eqref{eq:drag,eff}, NOVA-K
is employed to perform bounce averaged calculations since the bounce
motion is much faster than the perturbative mode evolution. Then,
a phase-space average needs to be performed to account for the contribution
of resonance surfaces spread over phase space, as described below.

\subsubsection{Bounce averaging}

The period of a particle poloidal bounce or transit motion is given
by $\tau_{b}=\oint dt=\oint d\theta r/\left({\bf v}_{\parallel}+{\bf v}_{d}\right)\cdot\hat{\theta}$,
where the drift velocity ${\bf v}_{d}$ is due to curvature and $\nabla B$
terms. Then, the bounce average of the effective collisional frequencies
is given by 
\[
\left\langle ...\right\rangle =\frac{1}{\tau_{b}}\oint\frac{d\theta r\left(...\right)}{\left({\bf v}_{\parallel}+{\bf v}_{d}\right)\cdot\hat{\theta}}
\]
For each set $\left(\mathcal{E},P_{\varphi},\mu\right)$, the trajectories
are \textit{a priori} known, hence there is no additional need to
explicitly follow the particles motion within the code.

\subsubsection{Phase space averaging}

The average over phase space is taken along the surfaces over which
the resonance condition is satisfied, for different poloidal bounce
harmonics. The phase space volume elements are weighted in accord
to their relative contribution to the overall growth rate. Specifically,
we evaluate

\[
\overline{\left(...\right)}=\frac{\int(...)Qd\Gamma}{\int Qd\Gamma}
\]
where $Q=\frac{1}{\omega}\underset{j}{\sum}\left\langle q_{EP}{\bf e}\cdot{\bf v}\right\rangle ^{2}\frac{\partial f}{\partial I}\delta\left(\Omega_{j}\right)$
is the contribution to the growth rate, $\gamma_{L}$, from a given
phase space location that satifies $\Omega_{j}\left(\mathcal{E},P_{\varphi},\mu\right)=0$.
In the present study, the phase-space averages are taken over several
harmonics of a given mode. This averaging technique was previously
used to predict the TAE amplitude saturation in TFTR experiments \cite{Gorelenkov1999Saturation}.

\subsection{Study of the chirping prediction for eigenmodes }

\begin{figure}[H]
\begin{centering}
\includegraphics[scale=0.3]{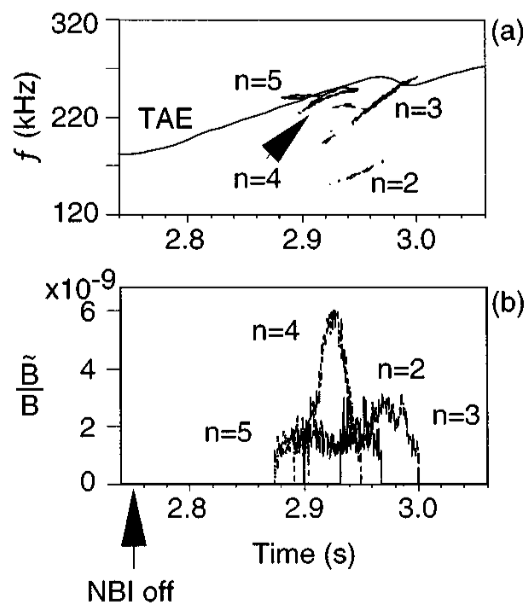}
\par\end{centering}
\caption{(a) Spectrogram of fixed-frequency, alpha-particle-driven TAEs in
TFTR and (b) relative amplitude of the TAEs with respect to the equilibrium
field for toroidal mode numbers $n=2-5$. Reproduced from \cite{FuNazikian1998}.\label{fig:TFTRSpectrogram}}
 
\end{figure}

\begin{figure}[H]
\begin{centering}
\includegraphics[scale=0.25]{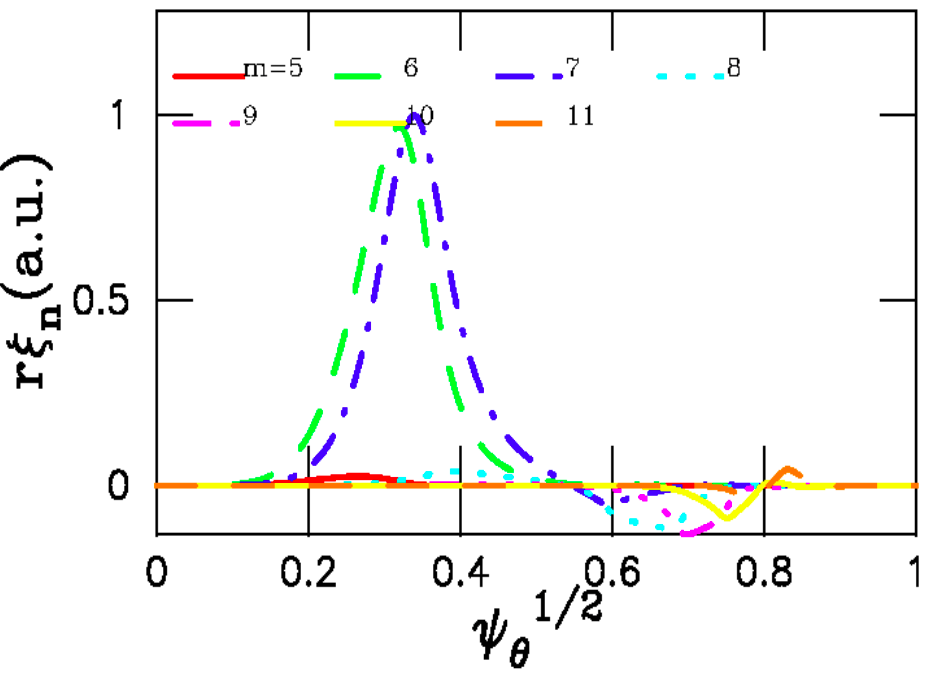}
\par\end{centering}
\caption{Mode structure of the dominant poloidal harmonics of a $n=4$ alpha-particle-driven,
core localized TAE in TFTR shot 103101 at $t=2.92s$. $\xi_{n}$ is
the fluid displacement and $\psi_{\theta}$ is the poloidal flux normalized
with its value at the plasma edge.\label{fig:StructureTFTR}}
 
\end{figure}

\begin{figure}[H]
\begin{centering}
\includegraphics[scale=0.3]{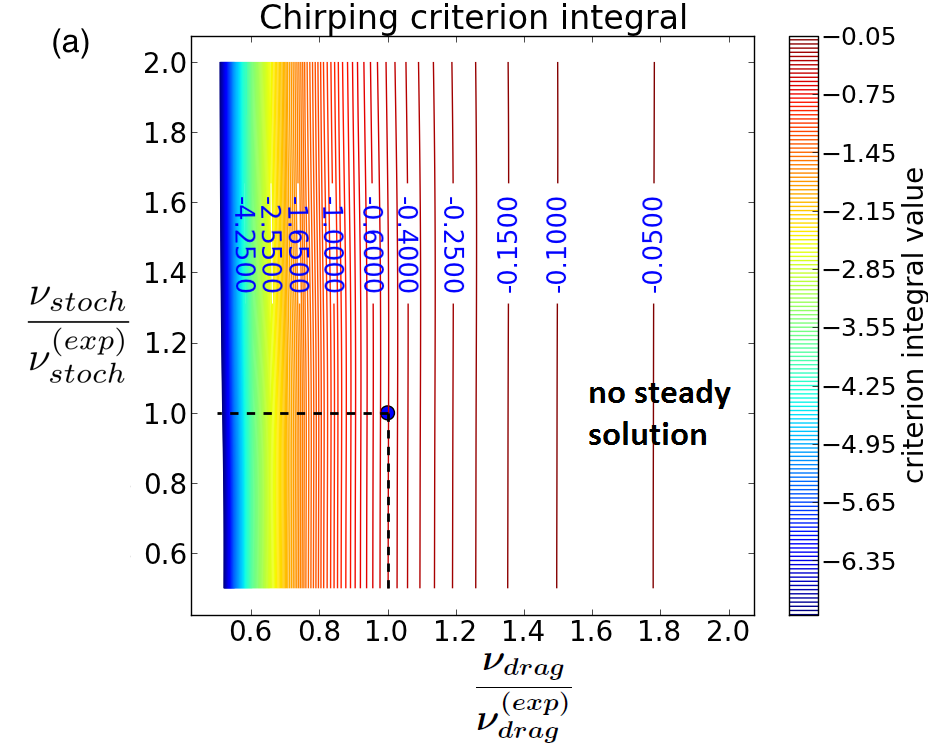}
\par\end{centering}
\begin{centering}
\includegraphics[scale=0.3]{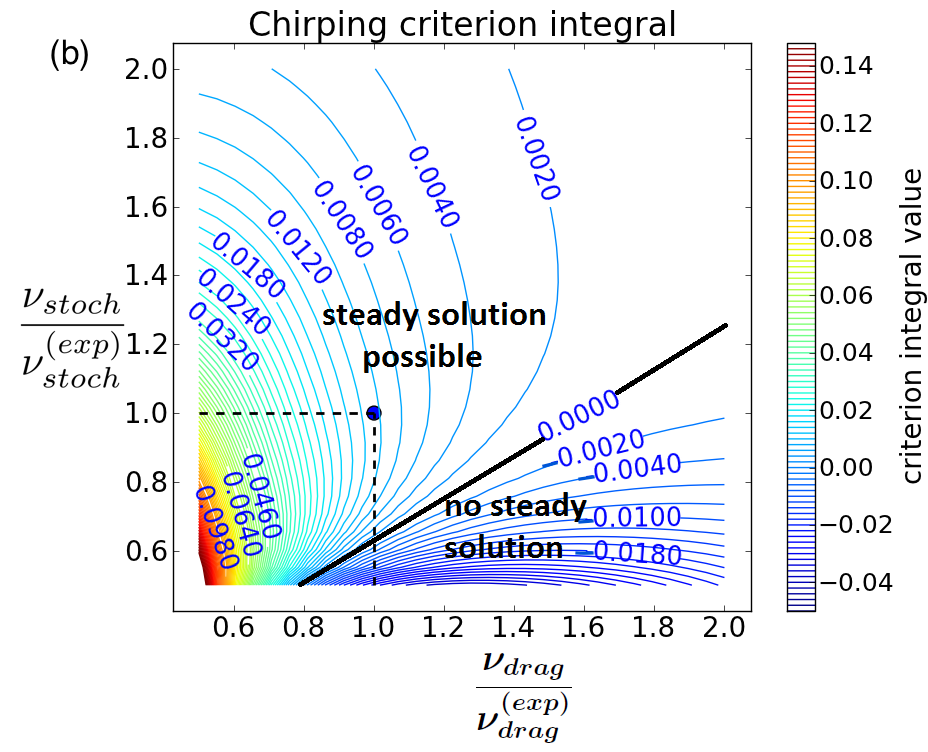}
\par\end{centering}
\caption{Plots showing contours of constant value of the criterion for chirping
existence, $Crt$, for an $n=4$ TAE in TFTR shot \#103101 (a) before
and (b) after the inclusion of micro-turbulence stochasticity. The
enhancement of stochasticity makes the point $\left(\nu_{stoch}/\nu_{stoch}^{(exp)}=1,\nu_{drag}/\nu_{drag}^{(exp)}=1\right)$
, that corresponds to the inferred experimental parameters, to cross
the boundary towards the positive region (where no chirping should
be allowed), which is consistent with the observation. Micro-turbulence
has a strong effect on bringing the mode to the steady state region,
thus suppresing the chirping. The contour plots are instructive in
order to analyz how far from the boundary (thick black line) the mode
is and therefore how prone it is to have its nonlinear character changed.
\label{fig:ContoursTFTR}}
 
\end{figure}

In order to show the importance of micro-turbulence as a chirping
suppression mechanism, an $n=4$ TAE mode driven by alpha particles
in the Tokamak Fusion Test Reactor (TFTR) shot 103101 was studied.
This mode, which was observed to oscillate at a constant frequency,
was analyzed in terms of the proposed criterion discussed above. Its
frequency and strength relative to the background field are presented
in Fig. \eqref{fig:TFTRSpectrogram}. The corresponding mode structure
obtained with NOVA code is shown in Fig. \eqref{fig:StructureTFTR}.
Shown in Fig. \eqref{fig:ContoursTFTR} are scans over the inferred
experimental values of $\nu_{stoch}$ and $\nu_{drag}$, denoted by
$\nu_{stoch}^{(exp)}$ and $\nu_{drag}^{(exp)}$, for the situations
(a) without the inclusion of micro-turbulent stochasticity and (b)
with its inclusion. 

A detailed visualization of how far the experiment is from the expected
boundary that separates the steady and chirping regions, as predicted
by the model, can be provided by scanning the criterion integral for
several values multiplying the actual $\nu_{drag}$ and $\nu_{stoch}$
(as shown in Fig. \eqref{fig:ContoursTFTR}). The vertical axis represents
the constants that are multiplying $\nu_{stoch}$ while the horizontal
one represents constants that are multiplying $\nu_{drag}$. The point
$\left(\nu_{stoch}/\nu_{stoch}^{(exp)}=1,\nu_{drag}/\nu_{drag}^{(exp)}=1\right)$
corresponds to the inferred experimental condition and the criterion
boundary is represented by the $0.0$ countour. Contours of negative
values represent regions in which chirping is expected while the positive
ones represent expected steady-state (fixed frequency). If experimental
conditions imply a positive $Crt$ sufficiently far from this transition
boundary, steady solutions should arise in experiment. 

Each contour plot corresponds to a given value of $Crt$ (eq. \eqref{eq:Criterion-1}),
as labeled in the two plots in Fig. \eqref{fig:ContoursTFTR}. In
part (a), where $Crt$ is evaluated only in terms of collisional pitch-angle
scattering and drag, $Crt$ is a negative number. Upon the addition
of micro-turbulent scattering, the re-evaluation of $Crt$ shows that
the point $(1,1)$ indicating the experimental conditions now operates
where $Crt>0$, which means that the criterion predicts that the mode
should not be able to chirp, being in agreement with the observation.
For this case, the estimation for $D_{EP}$ based on the scaling of
\cite{ZhangLinChen2008PRL} led to $D_{EP}\approx0.1m^{2}/s$, which
is consistent with direct measurements using beam blips in similar
experiments carried out in TFTR \cite{HeidbrinkSadler1994}. Numerically
calculated $\nu_{scatt}^{(exp)}$ and $\nu_{drag}^{(exp)}$ were $8568s^{-1}$
and $3138s^{-1}$, respectively. $\nu_{stoch}^{(exp)}$ was found
to be nearly $10$ times larger than $\nu_{scatt}^{(exp)}$ for this
particular mode. The $90^{\circ}$ degree pitch-angle scattering and
the inverse slowing down time were found to be $\nu_{\perp}=0.19s^{-1}$
and $\tau_{s}^{-1}=4.55s^{-1}$.

The NOVA-K code does not account for turbulence stochasticity, and
consequently it evaluates the criterion via an indirect method, using
the following procedure. First, we note that $Crt$ only depends on
the ratio of the overall stochasticity and collisional drag. From
equations \eqref{eq:nu_perp}, \eqref{eq:1/tau_S}, \eqref{eq:scatt,eff}
and \eqref{eq:drag,eff}, we see that the effective drag coefficient
is proportional to $T_{e}^{-3/2}$ while the effective collisional
scattering is insensitive to $T_{e}$. It is then possible to mimic
the inclusion of turbulence into $\nu_{stoch}$ by artificially decreasing
drag, which can be achieved by multiplying $T_{e}$ by a corresponding
numerical factor that leads to the correct value of the averaged values
of $\nu_{stoch}/\nu_{drag}$.

\begin{figure}[H]
\begin{centering}
\includegraphics[scale=0.32]{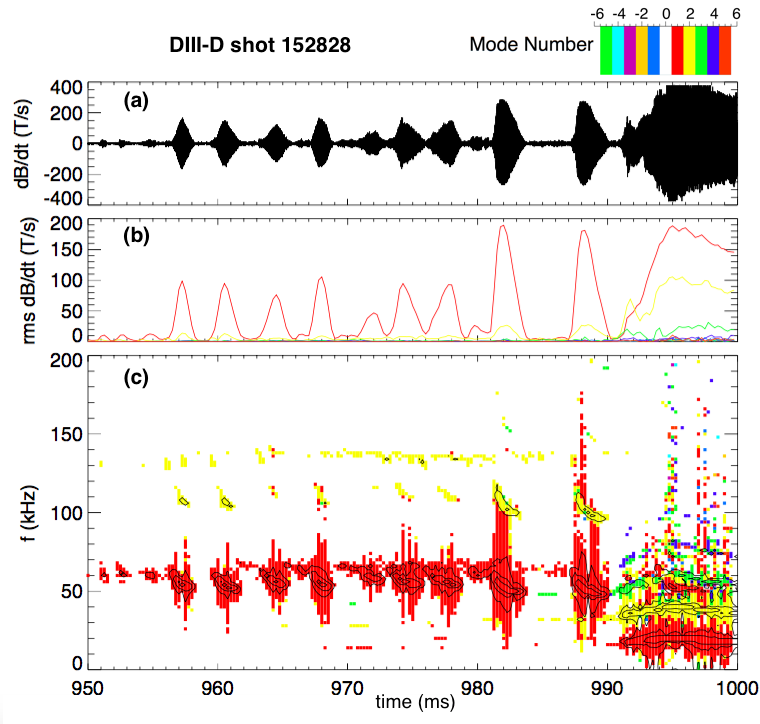}
\par\end{centering}
\caption{(a) time variation of the spectrum magnetic field $dB/dt$ for DIII-D
shot 152828, (b) root mean square of $dB/dt$ during and after the
chirping and (c) spectrogram showing the chirping fundamental and
second harmonics that exist before strong neoclassical tearing mode
starts at around $t=992ms$. \label{fig:AdditionalInfo28}}
 
\end{figure}

\begin{figure}[H]
\begin{centering}
\includegraphics[scale=0.6]{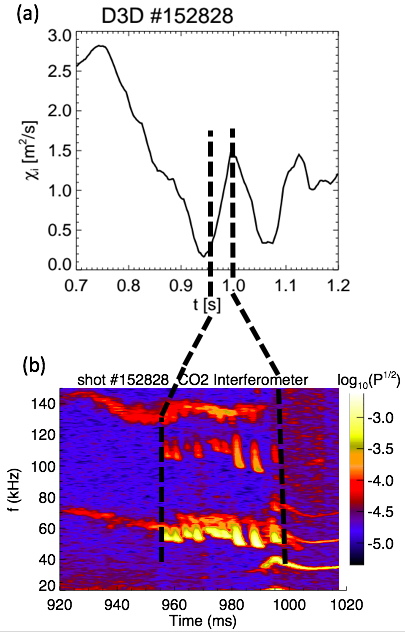}
\par\end{centering}
\caption{(a) Thermal ion conductivity and (b) spectrogram for DIII-D shot \#152828
as the mode undergoes transition from fixed-frequency to the chirping
regime. The chirping time window is indicated by the bold black dashed
lines. \label{fig:SpectrogramD3D}}
 
\end{figure}

\selectlanguage{american}%
A series of dedicated shots were performed on DIII-D with the objective
of triggering chirping and studying what conditions most strongly
determine a mode nonlinear evolution into the chirping regime. These
shots used high ion temperature at the core ($10-12keV$), $q_{min}$
$\sim2$ and strong toroidal rotation (up to $50kHz$ on axis). A
good example is the chirping observed in shot 152828, shown in Figs.
\eqref{fig:AdditionalInfo28} and \eqref{fig:SpectrogramD3D}.

\begin{figure}[H]
\begin{centering}
\includegraphics[scale=0.32]{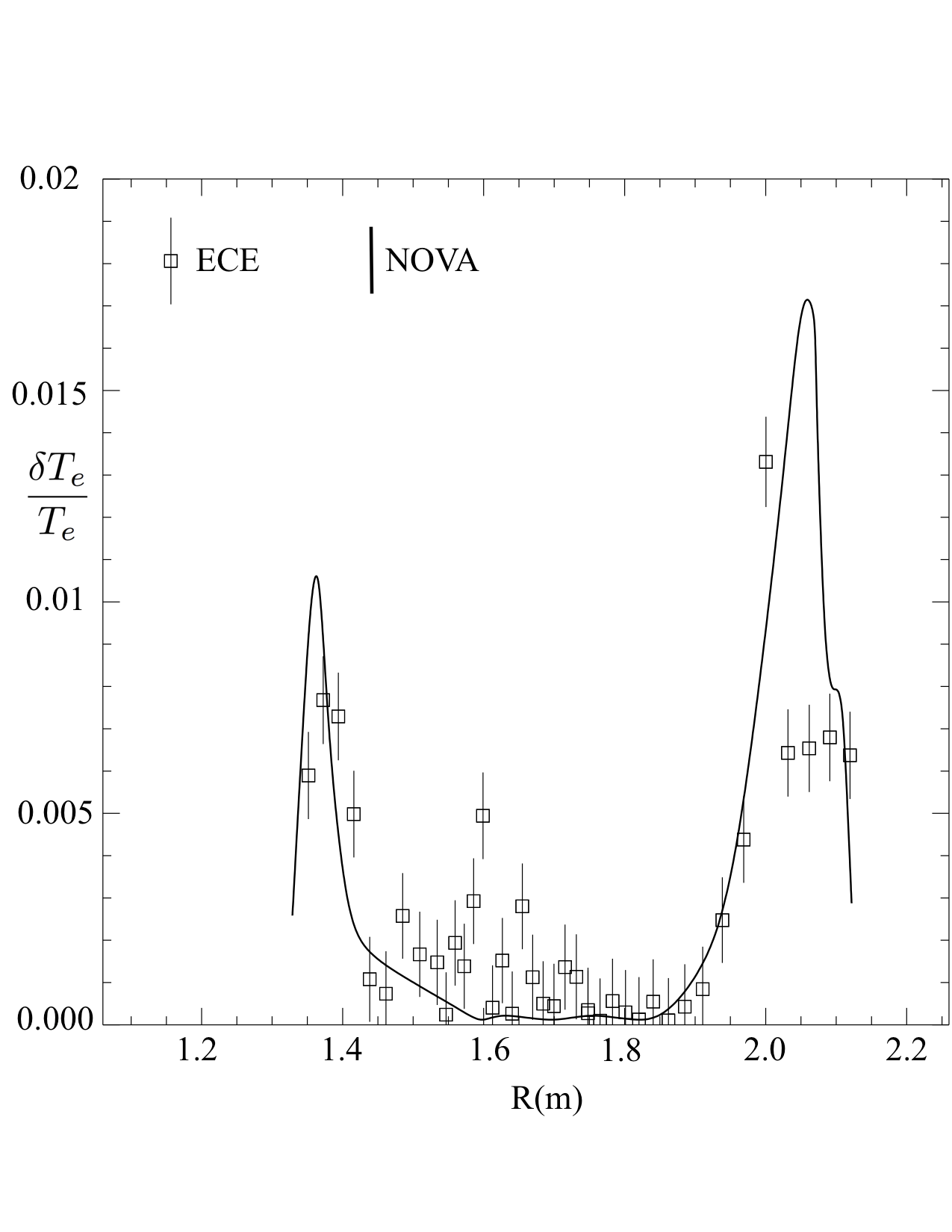}
\par\end{centering}
\caption{Electron temperature fluctuation reconstructed from NOVA $n=1$ BAAE
mode structure for DIII-D shot 152828 at $t=0.97s$ and its comparison
with electron cyclotron emission (ECE) measurements. The peak fluctuation
at the high field side, at around $R=1.4m$ is mostly due to fluid
compressibility while the peak at the low field side, close to $R=2.0m$
is mostly due to the fluid displacement itself, projected onto the
temperature gradient. \label{fig:DIIIDModeStructs}}
 
\end{figure}

The frequency of the chirping mode was too low to be a TAE. At first,
NOVA was run in a mode that enabled only the Alfvénic branch to be
captured and no reasonable mode structure was found. Then, NOVA was
run allowing both the Alfvénic and the sonic branches. A mode that
matches experimental evidence was obtained and identified to have
the characteristics of a beta-induced Alfvén-acoustic eigenmode (BAAE)
\cite{Gorelenkov2007BAAE} and is shown in Fig. \ref{fig:DIIIDModeStructs}.
The mode was identified using the data available on mode localization
from the electron cyclotron emission (ECE), frequency, local rotation
and toroidal mode number ($n=1$ for the chirping mode). For the comparison
with ECE data for electron temperature fluctuation, the fluid displacement
$\bm{\xi}$ calculated by NOVA was post-processed using the relation 

\[
\frac{\delta T_{e}}{T_{e}}=-\left(\gamma-1\right)\nabla\cdot\bm{\xi}-\frac{\nabla T_{e}}{T_{e}}\cdot\bm{\xi}
\]

\selectlanguage{english}%

\selectlanguage{american}%
The time-delayed cubic equation \eqref{eq:cubic-1} is derived assuming
that the nonlinear bounce frequency $\omega_{b}$ is much less than
the growth rate in absence of dissipation. This implies a limitation
on the cubic equation, with its domain of validity being restricted
to the early nonlinear phase, when mode amplitude, represented by
$\omega_{b}$, is still small compared to $\gamma_{L}$. From DIII-D
observations, we have noted that the chirping behavior typically starts
when $\chi_{i}$ drops to values lower than $0.25m^{2}/s$ \cite{DuarteAxivPRL}
but chirping persists even when $\chi_{i}$ raises to values that
would not admit the onset of the chirping process, according to the
theory. The chirping cycle appears to involve some degree of hysteresis
since once the first chirp happens, there is a tendency of the system
to continue chirping. Thus this observation appears to indicate that
once the chirping structures are already embedded in the system, continued
new chirping can still arise even though the apparent diffusivity
has increased to the point that chirping would not occur if phase
space structures were not already present. This hysteresis can be
noted in Fig. \eqref{fig:SpectrogramD3D} as the chirping persists
up to the point where $\chi_{i}\approx1.5m^{2}/s$.

Chirping does not restart when $\chi_{i}$ drops for the second time,
at $t=1.05s$. This is probably because of the effect of strong neoclassical
tearing modes (NTM) that are present at this time (as can be seen
from the parallel horizontal spectral lines that begin at around $t=992ms$
in Fig. \eqref{fig:SpectrogramD3D}). The suppression then probably
occurs because the NTMs are detuning the EPs from the resonance and
therefore contributes to extinguishing the drive.

\begin{figure}[H]
\begin{centering}
\includegraphics[scale=0.3]{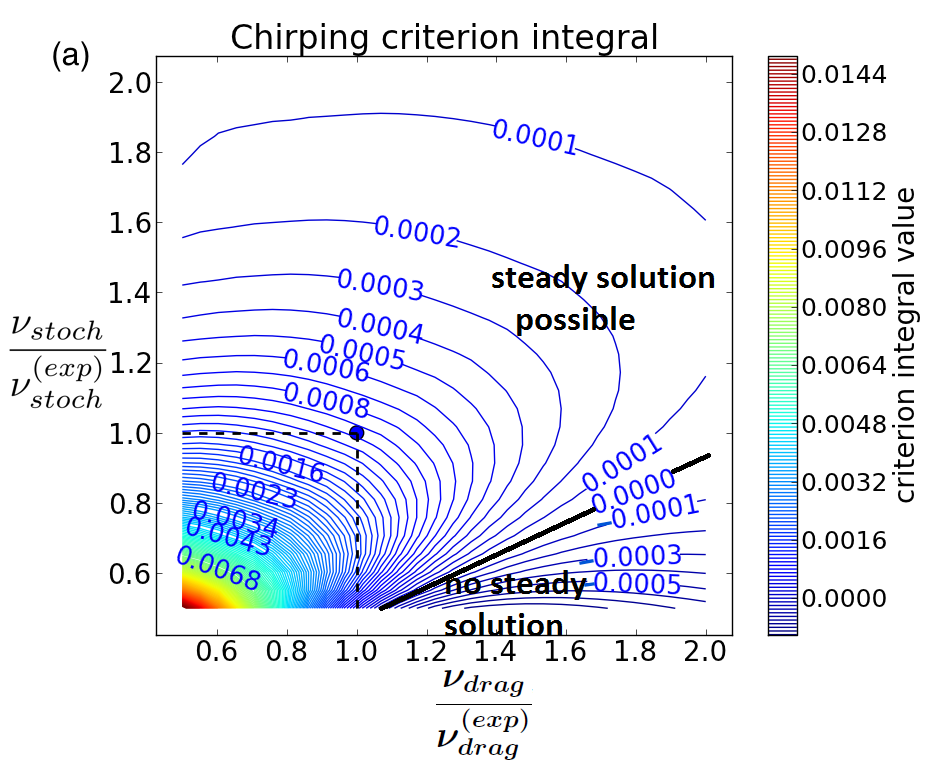}
\par\end{centering}
\begin{centering}
\includegraphics[scale=0.3]{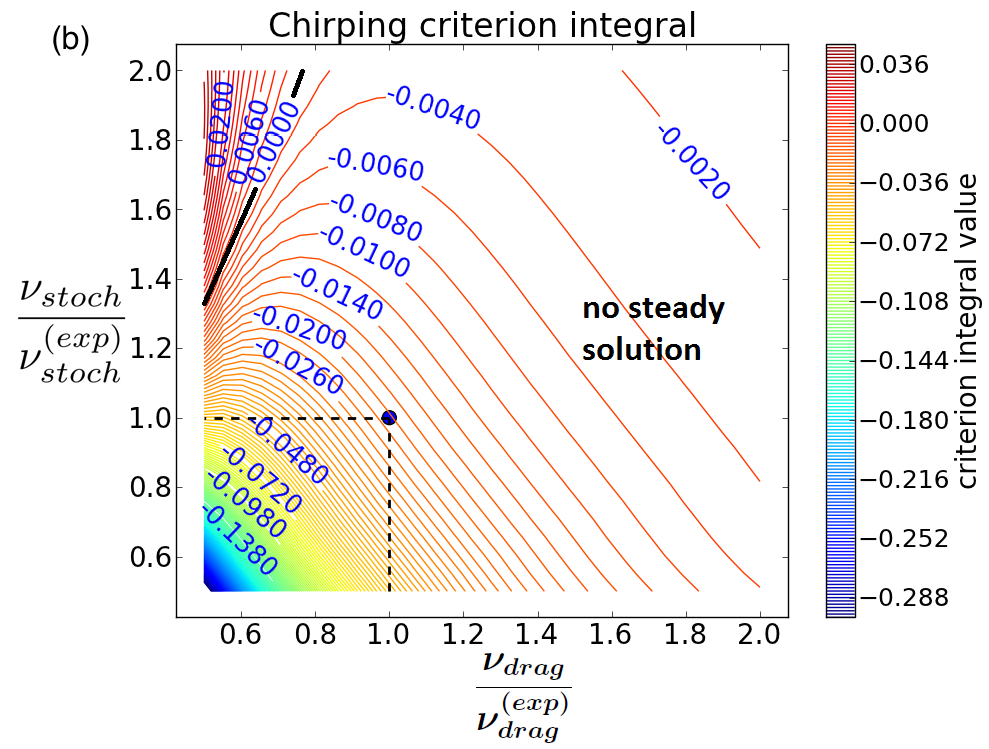}
\par\end{centering}
\caption{Plots showing contours of constant value of the criterion for chirping
existence, $Crt$, for an $n=1$ BAAE in DIII-D shot \#152828 shown
on the spectrogram of Fig. \eqref{fig:SpectrogramD3D} . The point
$\left(\nu_{stoch}/\nu_{stoch}^{(exp)}=1,\nu_{drag}/\nu_{drag}^{(exp)}=1\right)$
corresponds to the inferred experimental situation. (a) Before chirping
starts (at $t=900ms$, when $D_{th,i}\approx0.9m^{2}/s$) the criterion
integral $Crt$ is positive and (b) during chirping (at $t=960ms$,
when $D_{th,i}\approx0.25m^{2}/s$) the criterion integral is negative.
The point crosses the steady/chirping boundary represented by the
thick black line and has its nonlinear behavior changed. \label{fig:ContoursD3D}}
 
\end{figure}

\selectlanguage{english}%
Fig \eqref{fig:ContoursD3D} (a) shows the experimental condition
for the mode in DIII-D shot 152828 shown in Fig. \eqref{fig:DIIIDModeStructs}
before chirping starts (at $t=900ms$, when $D_{th,i}\approx0.9m^{2}/s$)
and Fig. \eqref{fig:ContoursD3D} (b) shows it during chirping (at
$t=960ms$, when $D_{th,i}\approx0.25m^{2}/s$), when the point has
transitioned from the positive to the negative region, in agreement
with the generalized criterion prediction. The criterion evaluated
close to the onset of chirping, at $t=960ms$, used the mode structure
calculated at $t=970ms$ (Fig. \ref{fig:DIIIDModeStructs}) since
trustworthy equilibria and profiles could not be made exactly at the
time of its onset.

\begin{figure}[H]
\begin{centering}
\includegraphics[scale=0.45]{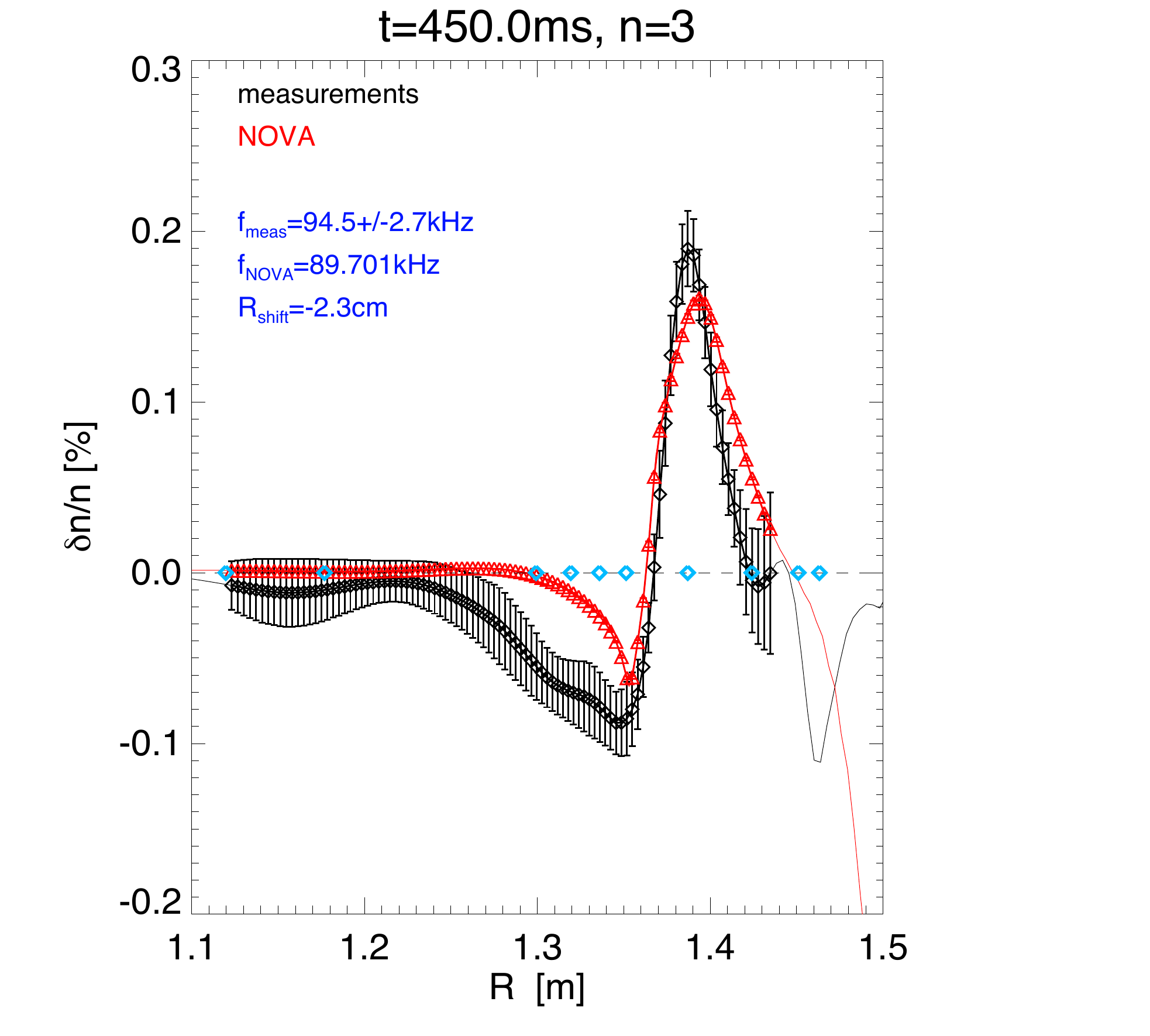}
\par\end{centering}
\caption{Comparison between the mode structure (quantified by the ratio of
the perturbed density $\delta n$ and the background density $n$)
inferred by the reflectometer (in black, with actual measurement locations
shown by the cyan diamonds) and the one calculated by NOVA code (in
red) for an $n=3$ TAE in NSTX shot 141711 at $t=450ms$.\label{fig:CompNOVAReflNSTX}}
 
\end{figure}

\begin{figure}[H]
\begin{centering}
\includegraphics[scale=0.3]{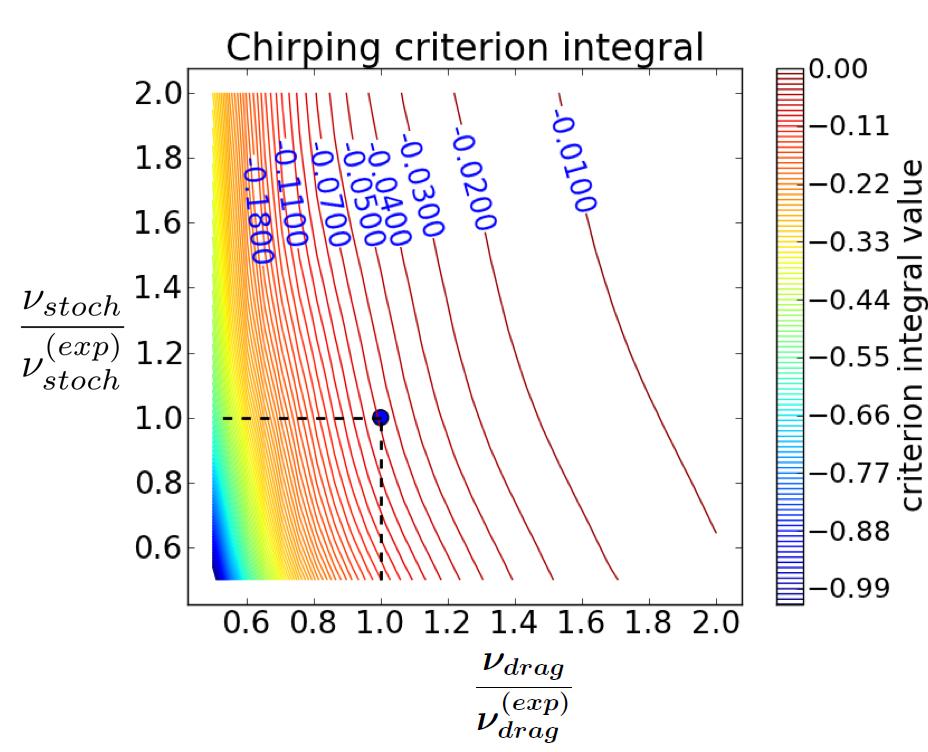}
\par\end{centering}
\caption{Plots showing contours of constant value of the criterion for chirping
existence, $Crt$, for an $n=3$ TAE in NSTX shot \#141711 at $t=450ms$
shown in Figs. \eqref{fig:SpectrogramNSTX} and \eqref{fig:CompNOVAReflNSTX}.
The point $\left(\nu_{stoch}/\nu_{stoch}^{(exp)}=1,\nu_{drag}/\nu_{drag}^{(exp)}=1\right)$
corresponds to the inferred experimental situation. .\label{fig:ContoursNSTX}}
 
\end{figure}

Similar framework of analysis was also applied to NSTX shot 141711,
whose spectrogram (Fig. \eqref{fig:SpectrogramNSTX}) showed chirping
for several TAEs with different toroidal mode numbers. A number of
chirping modes were studied, with the evaluation of $Crt$ for all
of them leading to negative numbers. A representative example is a
TAE that was catalogued by comparing the mode structures calculated
by NOVA with the reflectometer measurements, as shown in Fig. \eqref{fig:CompNOVAReflNSTX}.
The contour plot for $Crt$ for this mode is presented in Fig. \eqref{fig:ContoursNSTX}.
During the chirping time window in NSTX shot 141711, the ion transport
was dominated by neoclassical processes. It was observed that micro-turbulence
was not strong enough to suppress the bucket structures that sustain
chirping.


\section{Summary and discussions}

We have performed a study of the early phase of chirping events in
tokamak plasmas by means of realistic calculations of eigenstructures
and collisional coefficients. The proposed methodology has been described
in fair detail. In order to generalize the theoretical predictions
from the previous bump-on-tail approach, we have employed an action-angle
formalism for the general problem, with a similar perturbative procedure
used in \cite{BerkPPR1997}. It leads to the generalized criterion
for the likelihood of chirping and fixed-frequency solutions. It should
be noted that it only allows a steady solution to exist but does not
guarantee that the system will necessarily evolve to such a state.
It may happen that the evolution to steady states be only accessible
when higher-order nonlinear terms are taken into account, which are
not captured by the lowest nonlinear order perturbative approach used
by the framework of the time-delayed cubic equation (eq. \eqref{eq:cubic-1}). 

An interesting effect unveiled by the methodology used in this paper
is that the disparity of magnitudes of the positive and negative regions
of $Int$ (equation \eqref{eq:Int-1}), typically leads to the prediction
that a given mode should chirp, even for $\left\langle \nu_{stoch}/\nu_{drag}\right\rangle >1$,
unless a strong diffusive mechanism implies $\left\langle \nu_{stoch}/\nu_{drag}\right\rangle \gg1$.
This additional mechanism is observed to be often related to turbulent
processes. The contribution from regions of low $\mu$ turns out to
be quite important since the magnitude of $Int$ is larger for $\nu_{stoch}/\nu_{drag}\ll1$.
This shows the importance of using a weighted average of the transport
coefficients and not only a single representative point.

The employed approach is applicable for the modes near the stability
threshold. We implement it perturbatively using the mode structures
computed by the ideal MHD code NOVA. The perturbative approach is
justified when the growth rates evaluated for TAEs and RSAEs in NSTX,
DIII-D and TFTR plasmas are typically much smaller then the eigenfrequencies.
The case of unstable BAAEs requires more careful stability analysis
that is beyond the scope of this paper. Nevertheless observations
of those modes in DIII-D shown in Fig. \foreignlanguage{american}{\eqref{fig:SpectrogramD3D}(b)}
indicate that these modes are slowly growing, suggesting that they
are near the instability threshold.

We have set up plausible rules for determining the effective diffusivity
and drag. The main uncertainty in the diffusivity chosen for the EPs,
is the value of the diffusivity contribution from the background fluctuations.
We have chosen to base this value to be in accordance with what TRANSP
predicts to be the diffusion needed to produce the observed energy
confinement time. However, an additional uncertainty still remains
in determining how this thermal diffusion coefficient would be related
to the effective diffusion coefficient for the large orbit EP. We
chose to use an empirical formula obtained in ref. \cite{ZhangLinChen2008PRL},
based on electrostatic turbulence. However, the value chosen is rough,
and the actual scaling for a given experiment might differ considerably
from our choice. Using the estimates for EP diffusivity, we brought
in the model the effect of background turbulence, that adds up to
collisional scattering and contribute to suppress chirping. 

A number of factors that may influence the chirping formation are
not captured by the current theory. For example, static 3D fields
\cite{BortolonPRL2013} have been shown to affect bursting Alfvén
modes and reduce chirping. Besides that, effects such as toroidal
field ripples, energy diffusion, radio frequency heating fields \cite{MaslovskyMauelPRL2003,Heidbrink2006},
electromagnetic turbulence and neoclassical tearing modes can be important
in some scenarios to prevent chirping formation due to randomization
of phase space, with consequent resonance detuning. Other limitations
are that the cubic equation assumes small mode amplitude, which is
not necessarily the case in the experiment, and also that no mode
overlap has taken place.

Chirping events require self-trapped resonant particles to remain
locked with the excited chirping frequency for successive wave-trapping
bounce times. The maintenance of this time-dependent resonance condition
can induce significant convective transport over an extended region
of phase space. The present work can be helpful in addressing the
likelihood of convective and diffusive transport of fast ions and
therefore, be useful as a predictive tool for present-day and next-generation
devices.

\appendix

\section{Chirping likelihood in terms of the beam injection speed}

\selectlanguage{american}%
In planning and interpreting experiments, it can be meaningful to
understand the likelihood of observing wave chirping in terms of the
resonant particle speed. This is because the injected NB ions can
be either supra- or sub-Alfvénic which, for the case of TAEs, would
lead to dominant resonances located around $v_{\parallel}=v_{A}$
and $v_{\parallel}=v_{A}/3$, respectively. \foreignlanguage{english}{In
particular, a puzzling observation in JT-60U is that the abrupt large
events (ALEs) and their associated bursts and chirps usually happen
when the beam is injected supra-Alfvénically, using negative-ion-based
neutral beams, while fixed-frequency, quasi-steady modes are in general
observed when the NBI is sub-Alfvénic \cite{Kusama1999,Kramer2000,Shinohara2002,ShinoharaPPCF2004,Ishikawa2005}.
The ALEs have been intensively studied over the past decade \cite{BriguglioALEPoP2007,BierwageNF2013,BierwageNF2014,BierwagePoP2016a,BierwagePoP2016b,BierwageNF2017}.
Here we propose an explanation for the likelihood of these events
in terms of the framework adopted in this paper and we show that it
qualitatively agrees with JT-60U observations.}

Let us analyze a TAE for which the resonance condition is $\Omega=\omega_{TAE}+nv_{\parallel}/R-jv_{\parallel}/qR=0$\foreignlanguage{english}{.
For scenarios in which micro-turbulence can be ignored in comparison
with collisional scattering, the relevant ratio for the criterion,
eq. \eqref{eq:Criterion-1}, would be simply $\left(\nu_{scatt}/\nu_{drag}\right)^{3}$,
with }the effective scattering given by\foreignlanguage{english}{
eq. \eqref{eq:scatt,eff},}

\begin{figure}[H]
\centering{}\includegraphics[scale=0.4]{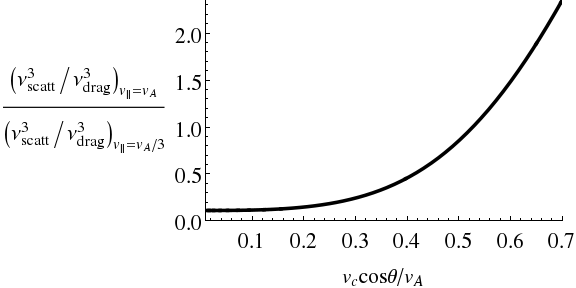}\caption{Ratio between the chirping-criterion-relevant parameter $\left(\nu_{scatt}/\nu_{drag}\right)^{3}$
for the cases in which the main resonance of a TAE is at $v_{\parallel}=v_{A}$
and $v_{\parallel}=v_{A}/3$, as a function of $v_{c}cos\theta/v_{A}$.\label{fig:Appendix}}
\end{figure}

\selectlanguage{english}%
\[
\nu_{scatt}^{3}\simeq\nu_{\perp}R^{2}v_{\perp}^{2}\left(\frac{jv_{\parallel}}{\omega_{c}qRr}\frac{dq}{dr}\right)^{2}
\]
where $\omega_{c}$ is the EP cyclotron frequency and \foreignlanguage{american}{$dP_{\varphi}\backsimeq-\left(\omega_{c}/q\right)rdr$}.
The effective drag is given by \eqref{eq:drag,eff},
\[
\nu_{drag}^{2}\simeq\frac{jRv_{\parallel}^{2}}{\tau_{s}\omega_{c}qr}\frac{dq}{dr}\left(1+\frac{v_{c}^{3}}{v^{3}}\right)
\]
with $v_{c}$ given by \eqref{eq:v_crit}. Therefore
\begin{eqnarray}
\frac{\nu_{scatt}^{3}}{\nu_{drag}^{3}} & = & \frac{v_{0}^{2}}{v^{2}}\frac{sin^{2}\theta/cos\theta}{\left(1+\frac{v_{c}^{3}}{v^{3}}\right)^{3/2}}\label{eq:RatioJT-60U}
\end{eqnarray}
where 
\[
v_{0}^{2}\equiv\frac{1}{2}\frac{\overline{A}_{i}}{\left[Z\right]}\frac{\left\langle Z\right\rangle }{A_{EP}}v_{c}^{3}\left(\frac{j\tau_{s}}{\omega_{c}qr}\frac{dq}{dr}\right)^{1/2}
\]
and $\theta$ is the angle between $\mathbf{v}$ and $\mathbf{B}$.
The ratio between the chirping-criterion-relevant parameter $\left(\nu_{scatt}/\nu_{drag}\right)^{3}$
prescribed by eq. \eqref{eq:RatioJT-60U}, for the cases in which
the main resonance of a TAE is at $v_{\parallel}=v_{A}$ and $v_{\parallel}=v_{A}/3$
is plotted in Fig. \eqref{fig:Appendix}. The ratio $\left(\nu_{scatt}/\nu_{drag}\right)_{v_{\parallel}=v_{A}}^{3}/\left(\nu_{scatt}/\nu_{drag}\right)_{v_{\parallel}=v_{A}/3}^{3}$
is $1/9$ for $v_{c}/v_{A}\approx0$. If $v_{c}cos\theta/v_{A}\lesssim0.5$,
it means that supra-Alfvénic injection implies more likelihood for
chirping. It is a direct implication of the chirping criterion (eq.
\eqref{eq:Criterion-1}) and the form of $Int$ (Fig. \eqref{eq:Int-1}).
Considering deuterium as the only background ion species, $n_{e}\approx n_{i}$,
$ln\Lambda_{i}\approx ln\Lambda_{e}$, and using typical parameters
for JT-60U, such as $B=3T$, $n_{i}=3\,10^{19}m^{-3}$ and $T_{e}=3keV$one
obtains $\frac{v_{c}}{v_{A}}\approx0.3$. The factor $cos\theta$
contributes to reduce the variable $v_{c}cos\theta/v_{A}$ used in
Fig. \eqref{fig:Appendix}. Therefore, because of the resonant velocity
dependence, if all other parameters are held fixed, supra-Alfvénic
injection implies more chances of observing wave chirping for TAEs
in JT-60U than in the sub-Alfvénic situation. Similar conclusion would
also hold for other conventional tokamaks, such as DIII-D. %

The previous conclusion was based on the fact that the stochasticity
coming from micro-turbulence is low compared to the one coming from
pitch-angle scattering.\foreignlanguage{american}{ However, the inclusion
of micro-turbulence using the scalings found in \cite{ZhangLinChen2008PRL}
does not change the conclusion that higher resonant velocities imply
more likelihood for chirping, provided that $v_{c}/v_{A}$ is sufficiently
small. In fact, turbulence even contributes to strenghten this interpretation.
This is because \cite{ZhangLinChen2008PRL} found that the micro-turbulence
diffusivity goes as $v^{-2}$ for passing particles and $v^{-4}$
for trapped particles, which is, similar to the effective pitch-angle
scattering, an inverse velocity dependence. Therefore micro-turbulence
contributes to further reduce the ratio }$\left(\nu_{scatt}/\nu_{drag}\right)_{v_{\parallel}=v_{A}}^{3}/\left(\nu_{scatt}/\nu_{drag}\right)_{v_{\parallel}=v_{A}/3}^{3}$,
which makes chirping for supra-Alfvénic injection even more probable
than in the purely collisional case.
\begin{acknowledgments}
We acknowledge fruitful discussions with G.-Y. Fu, E. D. Fredrickson
and R. M. O. Galvão. This work was supported by the São Paulo Research
Foundation (FAPESP, Brazil) under grants 2012/22830-2 and 2014/03289-4,
and by US Department of Energy (DOE) under contracts DE-AC02-09CH11466
and DE-FC02-04ER54698. This work was carried out under the auspices
of the University of São Paulo - Princeton University Partnership,
project ``\textit{Unveiling Efficient Ways to Relax Energetic Particle
Profiles due to Alfvénic Eigenmodes in Burning Plasmas}''.
\end{acknowledgments}

\end{document}